\documentclass[a4paper,showkeys,floatfix,aps,pre,longbibliography,superscriptaddress,reprint]{revtex4-2}
\usepackage{graphics,graphicx}
\usepackage{amsmath,amssymb}

\usepackage{graphics,graphicx}
\usepackage{dcolumn,bm}
\usepackage{psfrag}
\usepackage{xstring}
\usepackage{color}
\newcommand{\srm}
{\affiliation{Department of Physics, SRM University - AP, Amaravati,
 Andhra Pradesh - 522240, India}}

\begin{document}
\title{Exponent spectrum of Lorenz curves and its relation to system's heterogeneity}

\author{Soumyaditya Das}

\email{soumyaditya\_das@srmap.edu.in}
\srm

\author{Soumyajyoti Biswas}
\email{soumyajyoti.b@srmap.edu.in}
\srm

\begin{abstract}
    We analyze the effect of microscopic heterogeneity on the Lorenz curve of macroscopic observables. Lorenz curve of a response function being a cumulative and bounded quantity, is often a more stable function than the corresponding probability density. We show here that by doing an exponent spectrum analysis of the complementary Lorenz curve, it is possible to obtain a reflection of the underlying heterogeneity that causes the response function to depart from a power law behavior. We demonstrate this framework first by synthetic data and then by analyzing the avalanche statistics of a two dimensional, Random Field Ising Model (RFIM) at zero temperature. This method can lead to possible use in estimating microscopic heterogeneity of a system from analysis of an estimated Lorenz curve, particularly in socio-economic and physical contexts where the full probability distribution function is unavailable. 
\end{abstract}

\maketitle

\section{Introduction}
The Lorenz function was introduced in 1905 \cite{lorenz} to help quantify wealth inequality in a society. It is a cumulative measure, in the sense that $\mathcal{L}(p)$ quantifies the cumulative fraction of the wealth possessed by the poorest $p$ fraction of people in a society. From that one can then construct several inequality indices that help assign a number to the wealth inequality in a society. By construction, $\mathcal{L}(0)=0$ and $\mathcal{L}(1)=1$, and it is otherwise a monotonic, concave function. In the extreme limit of complete equality, the function is simply a diagonal straight line and in the other extreme where a single individual possesses all the wealth, it is $\mathcal{L}(p)=\delta_{p,1}$. 

Among the various inequality indices that can be extracted from the Lorenz function, the most widely known is the Gini index ($g$) \cite{gini}, which is the area between the diagonal line and the Lorenz curve, normalized by the area under the diagonal line (1/2). Other than that, the relatively new Kolkata index \cite{kolkata} is arguably the most intuitive one, which is simply the intersection of the Lorenz curve with the off-diagonal line (at $k,1-k$), which marks the fraction $k$ of the total wealth that is possessed by $1-k$ fraction of the population. It is then a generalization of the Pareto's 80-20 law \cite{pareto}, which appears as a special case when $p=0.8$. 

Over the years, just like the Pareto's 80-20 law, the Lorenz curve in general and the Gini index in particular, have found their uses in various other disciplines, including in inequality in other social resources (for example, in scientific citations \cite{neda,ghosh}), disease spreading \cite{wool} etc. More recently, inequality indices have also been studied for the responses of physical systems near their critical point, where those responses are known to be highly unequal \cite{succ}. It provides a remarkably simplified and universal representation of the critical phenomena and scaling exponents in terms of the Gini index \cite{prl}, in the sense that the critical scaling can then be written without the critical point of the system, which is a highly non-universal quantity. Furthermore, using the Kolkata index, a criterion for detecting an imminent large response could be formulated, which was then used in several numerical (see e.g., \cite{lomov}) and experimental (see e.g., \cite{jordi}) systems. Other than that in many cases of empirical data analysis and numerical simulations, it was shown that the inequality indices have near-universal behavior in many Self-Organized Critical (SOC) systems irrespective of their universality class \cite{manna}. 

In the above examples, the probability distribution of the response variable is usually known and the Lorenz curve is constructed from it, after which the inequality indices are calculated and analyzed. However, in many systems the accurate knowledge of the probability distribution could be unavailable, even though the grouped data at various hierarchy could still make it possible to have an accurate estimate of the cumulative measure such as the Lorenz curve \cite{econ1,econ2}. This is most common in the income and wealth distributions, where the individual level data is almost never globally accessible. However, various quantile data are often available and the Lorenz function can be estimated from it \cite{econ3,econ4}. Other than that in many physical systems, such as those showing crackling noise \cite{sethna}, the distribution data could be noisy, sparse or affected by binning. The Lorenz function, by construction, is far more stable and rarely affected by finite noise, (moderate level) sparsity of data or bin sizes \cite{newman}. 

Our aim in this work is to extract signatures of heterogeneity in the underlying system, by the analysis of the structure of the Lorenz curve. Specifically, we look at the exponent spectrum (or spectral density) of the (complementary) Lorenz curve when expressed in a polynomial basis. There has been an earlier effort to expand the Lorenz function using Legendre polynomial \cite{leg}, which is an orthogonal basis. But we use a simple power-law basis, so as to preserve the direct correspondence between the power and the corresponding exponent of the probability distribution. While a direct inversion to obtain the exponent spectrum, and hence the estimate of the exponent for the probability distribution, is ill-posed, several indirect ways lead to reliable estimates of the width of the exponent spectrum. These measures can then point to a quantitative estimate of the underlying heterogeneity, which often influence the responses of the systems. After building the framework, we test it first on synthetic data, for which the exponent spectra are known. Following this, we apply the method to Random Field Ising Model (RFIM) \cite{rfim1,rfim2,rfim3,rfim4,rfim5}, and show that near the critical point, the heterogeneity strength (width of the distribution of the random fields) is highly correlated with the (indirect measures of) spectral width of the Lorenz function.

\section{Results} 
\subsection{Formulation}
Let us consider a power law size distribution function in the form of $P(S)=CS^{-\delta}$, where $\delta >1$. Generally, for this type of distribution, a lower cut-off as well as a higher cut-off are required to make the mean of the distribution finite. Now, if the event sizes are arranged in ascending order with $m$ denoting the index of ordering), it will show a power law diverging function such as $S_m\sim (m-m_0)^{-n}$, where the divergence occurs at $m=m_0$, with $n=\frac{1}{\delta-1}$. It is then possible to estimate the Lorenz function ($\mathcal{L}(p)$) from the diverging function (up to $b=\frac{m}{m_0}$), which is given by (see \cite{prl}). 
\begin{equation}
    \mathcal{L}(p,b)=\frac{1-(1-pb)^{1-n}}{1-(1-b)^{1-n}}
\end{equation}

Note that above equation for $\mathcal{L}(p)$ is valid for any $n$ (for a given $b$) except $n=1$ and $n=2$.  However, for $b=1$, $n$ must be less than 1 (meaning $\delta > 2$ and in that case, the higher cut-off can be taken as infinity to get the mean of the distribution). Hence, the Lorenz function becomes 
\begin{equation}
    \mathcal{L}(p)=1-(1-p)^{1-n}
\end{equation}
Again, the above equation can be rewritten in terms of the complementary Lorenz function $\mathcal{L}^*(q)=1-\mathcal{L}(1-p)$, where $q=1-p$.

\begin{equation}
    \mathcal{L}^*(q)=q^\theta; \theta=1-n
\end{equation}
    This functional form of the $\mathcal{L}^*(q)$ comes due to the power-law diverging response function. Any departure from the power-law divergence would give a different functional form of $\mathcal{L}^*(q)$ which is usually very hard to estimate. Consider a function $\mathcal{B}^*(q)$ defined by the integral transform of the spectral density $A(\theta)$ as,
\begin{equation}
    \mathcal{B}^*(q)=\int_{0}^{\infty}A(\theta)q^\theta d\theta
\end{equation}
Where the kernel of the transformation is $q^\theta$ and a spectral density is valid only if $\int_{0}^{\infty}A(\theta) d\theta=1$. We can immediately see that if $A(\theta)=\delta(\theta-\theta_0)$, $\mathcal{B}^*(q)=q^{\theta_0}$, that is, we get back the complementary Lorenz function $\mathcal{L}^*(q)$. That also suggests a narrow spectral density ($\delta$-function) gives a broad size distribution (power law type).    

Now, let us consider a broad spectral density,
\begin{equation}
    A(\theta)=\frac{1}{\theta_2-\theta_1}\Theta(\theta-\theta_1)\Theta(\theta_2-\theta)
\end{equation}
where $\Theta(.)$ is a Heaviside step function. Again, $\mathcal{B}^*(q)$ can be calculated from Eq. 4. 

\begin{equation}
    \mathcal{B}^*(q)=\frac{q^{\theta_2}-q^{\theta_1}}{(\theta_2-\theta_1)\ln q}
\end{equation}
$q=1-p$ and $1-\mathcal{B}^*(p)=\mathcal{B}(p)$
\begin{equation}
    \mathcal{B}(p)=1-\frac{(1-p)^{\theta_2}-(1-p)^{\theta_1}}{(\theta_2-\theta_1)\ln (1-p)}
\end{equation}
Note that in general $\mathcal{B}(p)$ does not satisfy all the properties of the Lorenz function, unless $\theta_1,\theta_2<1$. Therefore, if we consider $\mathcal{B}(p)$ is a Lorenz function then we need to have $\theta_1,\theta_2<1$. For the uniform spectral density $A(\theta)$ we can expand it in terms of the power $\varepsilon$, $\theta_2=\theta_0+\varepsilon, \theta_1=\theta_0-\varepsilon$, where $\theta_0$ and  $2\varepsilon$ are mean and width of $A(\theta)$ respectively. 

Rewriting the above equation in terms of $\varepsilon$
\begin{equation}
     \mathcal{B}(p)=1-(1-p)^{\theta_0}\bigg[\frac{\sinh{[\varepsilon\ln(1-p)]}}{\varepsilon\ln(1-p)}\bigg]
\end{equation}
Expanding in Taylor series (for small $\varepsilon$) gives the form
\begin{equation}
    \mathcal{B}(p)=1-(1-p)^{\theta_0}\bigg[1+\frac{\varepsilon^2}{6}[\ln(1-p)]^2+O(\varepsilon^4)\bigg]
\end{equation}

Now we will calculate the probability density function (PDF) from $\mathcal{B}(p)$. 
\begin{equation}
    P\{x(p)\}=\frac{1}{\mu\mathcal{B}^{''}(p)} ; \\
    x(p)=\mu\mathcal{B}^{'}(p)
\end{equation}
From Eq. 9 we get the PDF,

\begin{widetext}
\begin{equation}
\begin{aligned}
P(x(p)) &= \frac{1}{{\mu (1-p)^{\theta_0 - 2}}\Bigg[\theta_0(1-\theta_0)+\frac{\varepsilon^2}{6}\theta_0(1-\theta_0)(\ln(1-p))^2 
+ \frac{\varepsilon^2}{3}(1-2\theta_0)\ln(1-p)
- \frac{\varepsilon^2}{3}
\Bigg]} \\
\\
x(p) &= \mu (1-p)^{\theta_0 - 1} \left[
\theta_0 
+ \frac{\varepsilon^2}{6} \left( \theta_0 (\ln(1-p))^2 + 2\ln(1-p) \right)
\right]
\end{aligned}
\end{equation}
\end{widetext}

 When $\varepsilon=0$, Eq. (11) reduces to simple power law decay type distribution,
\begin{equation}
    P(x)= \frac{1}{\mu\theta_0(1-\theta_0)}x^{-\left(\frac{\theta_0-2}{\theta_0-1}\right)}
\end{equation}
Where the exponent $\delta=\left(\frac{\theta_0-2}{\theta_0-1}\right)$. Here $\delta$ is always greater than 2.

\subsection{Case: $n>1$}
We discussed earlier that the above formalism is valid only for $n<1$. Here we shall discuss the case of $n>1$ (i.e., $1<\delta<2$) since it is of the most practical importance. Now we can make a transformation of $S_m$ (the diverging response function) such that it becomes, $\tilde{S}_m=(S_m)^r\sim (m-m_0)^{-\tilde{n}}$, where $\tilde{n}=nr$ and here although $n>1$ but $0<nr<1$. 

Hence, the same procedure can be repeated to obtain the Lorenz function $\mathcal{\tilde{B}}(p)$ and then the probability density function $\tilde{P}(\tilde{x})$. The Lorenz function has the form,
\begin{equation}
    \mathcal{\tilde{B}}(p)=1-(1-p)^{\tilde{\theta}_0}\bigg[1+\frac{\tilde{\varepsilon}^2}{6}[\ln(1-p)]^2+O(\tilde{\varepsilon}^4)\bigg]
\end{equation}
where $\tilde{\theta}_0=1-nr=\frac{\tilde{\theta}_1+\tilde{\theta}_2}{2}$ and $\tilde{\varepsilon}=\frac{\tilde{\theta}_2-\tilde{\theta}_1}{2}$.

Similarly, the probability density function can be written as

\begin{widetext}
\begin{equation}
\begin{aligned}
\tilde{P}(x(p)) &= \frac{1}{{\mu (1-p)^{\tilde{\theta}_0 - 2}}\Bigg[\tilde{\theta}_0(1-\tilde{\theta}_0)+\frac{\tilde{\varepsilon}^2}{6}\tilde{\theta}_0(1-\tilde{\theta}_0)(\ln(1-p))^2 
+ \frac{\tilde{\varepsilon}^2}{3}(1-2\tilde{\theta}_0)\ln(1-p)
- \frac{\tilde{\varepsilon}^2}{3}
\Bigg]} \\
\\
x(p) &= \mu (1-p)^{\tilde{\theta}_0 - 1} \left[
\tilde{\theta}_0 
+ \frac{\tilde{\varepsilon}^2}{6} \left( \tilde{\theta}_0 (\ln(1-p))^2 + 2\ln(1-p) \right)
\right]
\end{aligned}
\end{equation}
\end{widetext}
 
When $\tilde{\varepsilon}=0$, Eq. (14) reduces to simple power law decay type distribution,
\begin{equation}
    \tilde{P}(x)= \frac{1}{\mu\tilde{\theta}_0(1-\tilde{\theta}_0)}x^{-\left(\frac{\tilde{\theta}_0-2}{\tilde{\theta}_0-1}\right)}
\end{equation}
The above equation can also be written in terms of $\theta_0(=1-n)$, with $\theta_0=1-\frac{(1-\tilde{\theta}_0)}{r}$ and $(0<\tilde{\theta}_0<1)$
\begin{equation}
    \tilde{P}(x)\sim x^{-\left(\frac{\theta_0-2}{\theta_0-1}\right)}
\end{equation}
As $\theta_0$ is negative ($n>1$), the size distribution exponent $\delta=\left(\frac{\theta_0-2}{\theta_0-1}\right)$ lies between 1 and 2 (i.e., $1<\delta<2$).

\begin{figure}[tbh]
    \hspace*{-0.5cm}
    \includegraphics[width=1.\linewidth]{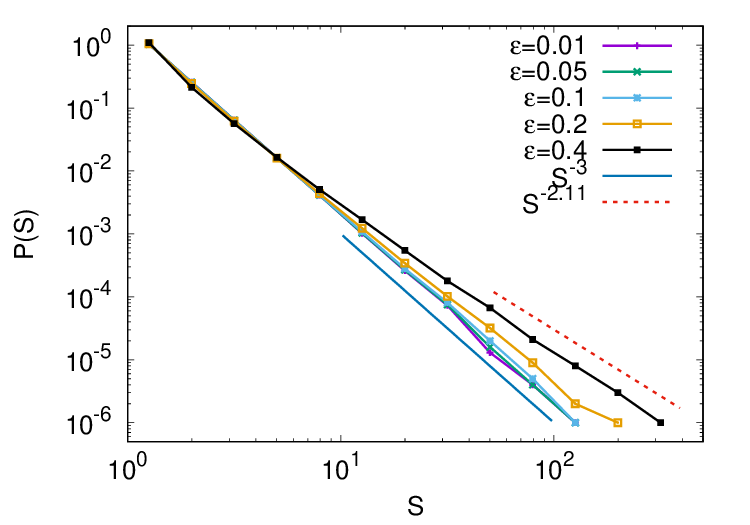}
    \caption{The probability distribution of the synthetic data generated from assuming a uniform spectral density $A(\theta)$ in the range ($\theta_0-e,\theta_0+e$), with $\theta_0=0.5$. For small values of $e$, $\delta\approx 3.0$ and for larger values of $e$, it is expected to show the slowest decay available, which for $e=0.4$ is $\delta\approx2.11$, as indicated.}
    \label{synth_dist}
\end{figure}
\begin{figure}[tbh]
    \hspace*{-1.cm}
    \includegraphics[width=1.\linewidth]{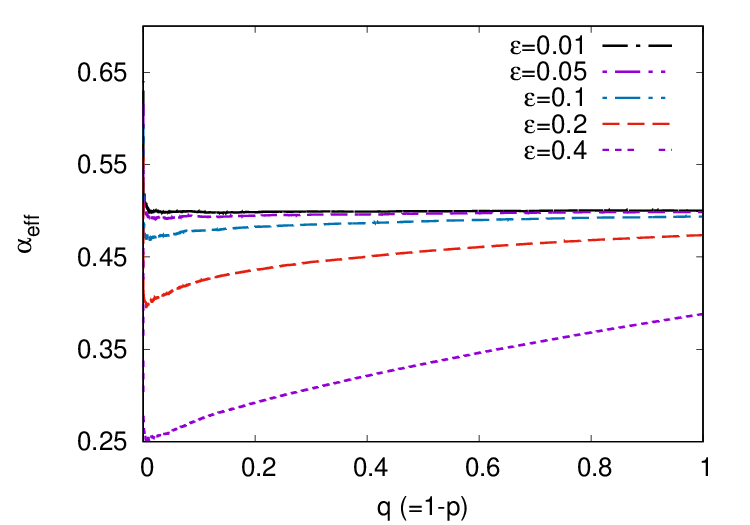}
    \caption{The variation of the effective exponent for synthetic data with $q$ is shown for different widths of the spectral density. As the spectral density is chosen to be centered at $\theta_0=0.5$, for small widths, the effective exponent remains close to that for all values of $q$. However, for larger width, the effective exponent starts deviating, especially towards the larger event side ($q\to 0$).}
    \label{synth_alpha}
\end{figure}
\subsection{Analysis with synthetic data}
Since the spectral density $A(\theta)$ is not a controllable function for a realistic system, it is useful to verify, as an intermediate step, how far can its effect be extracted from a Lorenz function. For this, we keep the form of the spectral density in Eq. (5), with $\theta_1=\theta_0-\varepsilon$ and $\theta_2=\theta_0+\varepsilon$, with $\theta_0=0.5$. For the synthetic data, the probability distribution is first generated as follows: randomly a $\theta$ is drawn from within the range ($\theta_1,\theta_2$). It is then converted to the size distribution exponent as outlined above: $\delta=\frac{\theta-2}{\theta-1}$. A random variable is then constructed using the inverse transform sampling $S=(1-u)^{-1/(\delta-1)}$, with $u\in [0,1]$. One can then obtain the corresponding probability distribution $P(S|\theta)\sim S^{-\delta(\theta)}$, which is essentially a uniform mixture of all exponents within the range ($\theta_1,\theta_2$). Fig. \ref{synth_dist} shows such distribution functions constructed for various values of $\varepsilon$ and $\theta_0=0.5$. Clearly, for small values of $\varepsilon$, one would expect $\delta=3$, as can be seen. For higher values of $\varepsilon$, the most slowly decaying exponent is expected to dominate, i.e. $\delta_{min}=2.11$ (for $\varepsilon=0.4$).

There are realistic circumstances where such a mixture of power laws can appear. It can happen, for example, in Pareto like wealth distribution among various sub-populations, avalanche sizes in heterogeneous samples \cite{two_power} etc. As we shall see, in RFIM, the width of the distribution of the random fields can affect the power law of the avalanche sizes (driven slowly by external magnetic field). Each of these scenario indicates, either explicitly (for RFIM) or implicitly (for wealth distribution) that underlying microscopic heterogeneity has a role to play in determining the shape of the distribution function and in turn that of the spectral density $A(\theta)$. 

In this case, from the probability density, we construct the Lorenz function ($\mathcal{L}(p)$) and its complement ($\mathcal{L}^*(q)$). For our choice of values for $\theta_0$ and $e$, we always have $\delta>2$, so no further variable transformation is required before constructing the Lorenz curves and its complement.  Here, by construction $\mathcal{L}^*(q)=\int\limits_{\theta_1}^{\theta_2}A(\theta)q^{\theta}d\theta$. However, in realistic situations, where the Lorenz function (and its complement) is known, doing the inverse transform to obtain the spectral density is an ill posed problem (as $q^{\theta}$ is not an orthogonal basis set). We then have to resort to estimating the spectral density through indirect means. We use those means here as well to benchmark their reliability in cases where $A(\theta)$ is not known a-priori. 

\begin{figure}[tbh]
    \hspace*{-0.5cm}
    \includegraphics[width=1.1\linewidth]{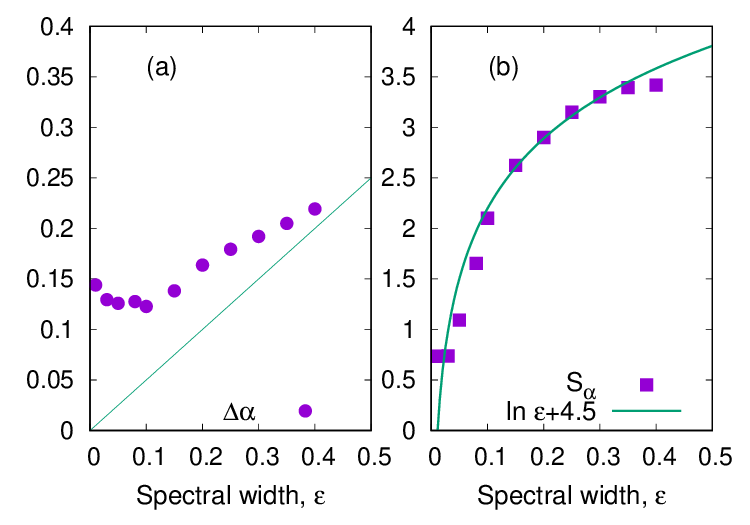}
    \caption{The variations of (a) $\Delta\alpha$ and (b) $S_{alpha}$ are plotted for different values of $e$. The first quantity is essentially the difference between the maximum and minimum local slopes of the complementary Lorenz function (see Eq. (\ref{alpha_width})), while the second one is the entropy of the distribution constructed from the values of the local exponents obtained (see Eq. (\ref{alpha_entropy})). The first quantity almost behaves linearly with $e$, indicating it could be taken as a representation of $A(\theta)$ (at least where it is uniform), when that is explicitly not available. It then means that the entropy of the distribution of $\alpha$ values could be associated with entropy of the spectral width, which in this case should vary as $\ln \varepsilon$, which seems to hold up to an additive constant.}
    \label{synth_e}
\end{figure}

We calculate the local exponent that describes the complementary Lorenz function for a particular value of $q$:
\begin{equation}
    \alpha_{eff}=\frac{d\ln \mathcal{L}^*(q)}{d\ln q}.
\end{equation}
In the case of a pure power-law distribution, one should have $A(\theta)=\delta(\theta-\theta_0)$, and consequently $\theta_0=\alpha_{eff}$. However, when there is a departure from the pure power-law distribution, $A(\theta)$ will have a finite width, as discussed above, and hence $\alpha_{eff}=\alpha_{eff}(q)$ will depend on $q$ (see Fig. \ref{synth_alpha}). The next step is then to construct measures that would faithfully mirror $A(\theta)$. We do this through two measures, 
\begin{equation}
    \Delta \alpha=(\alpha_{eff})_{max}-(\alpha_{eff})_{min},
    \label{alpha_width}
\end{equation}
and the spectral entropy
\begin{equation}
    S_{\alpha}=-\sum\limits_i Q(\alpha_{eff}^i)\ln Q(\alpha_{eff}^i),
    \label{alpha_entropy}
\end{equation}
where $Q(.)$ is the histogram of all $\alpha_{eff}$ values obtained for various values of $q$ (the sum also runs over all such values). Fig. \ref{synth_e} shows the variations of these two quantities with $e$. As can be seen, they are highly correlated with, hence a faithful mirror of the spectral width. It is possible to make a theoretical estimate of their functional forms. Of course, in analyzing the effective exponent $\alpha_{eff}$, we wish to capture the behavior of $\theta_0$, which in this synthetic case, is a good approximation. Now, as $A(\theta)=1/2\varepsilon$ in the range ($\theta_0-\varepsilon,\theta_0+\varepsilon$), the spectral entropy should simply be
\begin{equation}
    S_{\alpha}=-\int\limits_{\theta_1}^{\theta_2}A(\theta)\ln A(\theta)d\theta,
\end{equation}
which gives $S_{\alpha}\sim \ln \varepsilon$. We have retained the subscript $\alpha$, since in realistic cases (where $A(\theta)$ is not known), the correspondence $\alpha=\theta$ is only approximate. Also, in numerical calculations, we do a discrete sum, rather than a continuous integral, and the bin width appears as an additive constant in this relation. 
As for the range of the effective exponent $\Delta\alpha$, it should simply represent the width of $A(\theta)$ and that is $\varepsilon$ here. But given that is an extreme measure (difference between the maximum and minimum effective exponent values), fluctuations can severely affect its value. Therefore, in Fig. (\ref{synth_e}) we see a linear variation of $\Delta\alpha$ with $\varepsilon$, but it is closer to $\varepsilon/2$ (indicated by the solid line).

Therefore, we see using synthetic data that even though the direct evaluation of spectral width is difficult, we can have useful measures that we can apply on the Lorenz function and from which we have an idea of the spectral width. 

\subsection{Effect of disorder on spectral width of Random Field Ising Model avalanche distributions}
We now turn to a specific example of a physical system, where the spectral width can vary due to some physical property of the system. Specifically, we look at the $T=0$ limit of the Random Field Ising Model (RFIM) on a square lattice with periodic boundary conditions 
The Hamiltonian of the system reads
\begin{equation}
    H=-J\sum\limits_{\langle ij\rangle}s_is_j-\sum\limits_ih_is_i-H_{ext}\sum\limits_is_i,
\end{equation}
where the spin variables $s_i=\pm 1$, which interacts with the nearest neighbors (as indicated by the angular brackets in the limit of the sum) and $H_{ext}$ is the external magnetic field that is slowly tuned. 

The dynamics of the model proceed as follows: the spins are initially all aligned in one direction. Then the external field is slowly increased in the opposite direction. Due to the random (Gaussian distributed) fields ($h_i$), the spins requiring the lowest energy to flip will flip first. However, this changes the environment of the four nearest neighboring spins, some of which may again flip and so on. The value of the external field is held constant until there is no further spin flipping in the system. Due to the fact that $T=0$, this will be a stable state. Then the external field is again slightly ($dH=0.01$ here)
and the flipping dynamics may restart. This process continues until all spins have aligned in the direction of the external field. The number of spins flipped between two successive increase of the external field is considered an avalanche. 

\begin{figure}[tbh]
    \hspace*{-0.5cm}
    \includegraphics[width=1.\linewidth]{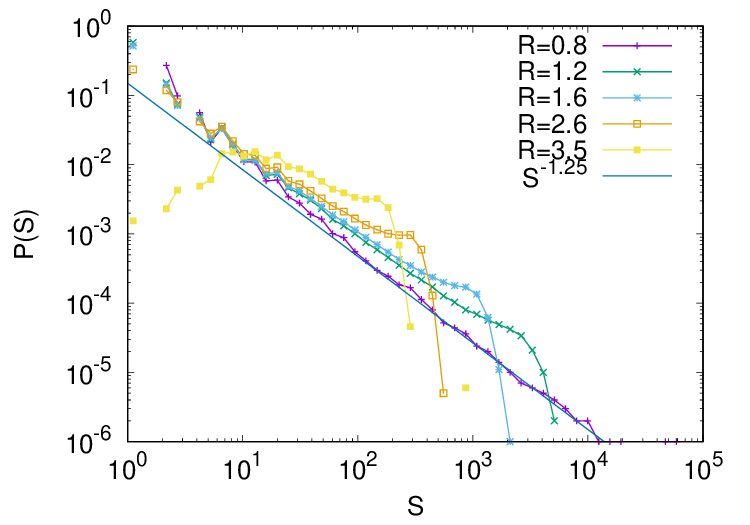}
    \caption{Avalanche size distribution of the 2d RFIM at $T=0$ for different values of the disorder strength. The power law is observed at critical $R_c=0.8$ and above that it deviates from power law. Below the critical disorder, all spins flip in one avalanche.}
    \label{rfim_avalanche}
\end{figure}

The random fields follow a distribution given by
\begin{equation}
    P(h_i)=\frac{1}{\sqrt{2\pi}R}\exp\left(-\frac{h_i^2}{2R^2}\right),
\end{equation}
where $R$ denotes the strength of the disorder. For the weak disorder case, due to increase in the external field, there will be a system spanning avalanche where all spins will reverse sign. On the other hand, when disorder is very wide, there will be only isolated spin reversals. There will be a critical value of $R=R_c\approx 0.8$ where the avalanche dynamics is scale free. Just above the critical disorder strength, there will be gradual deviation from the power law scaling, which should be captured by the spectral density analysis. We can then see the variations of the effective widths of spectral density and its relation to the underlying heterogeneity of the system, in this case the width of the random field distribution. 

\begin{figure}[tbh]
    \hspace*{-0.5cm}
    \includegraphics[width=1.\linewidth]{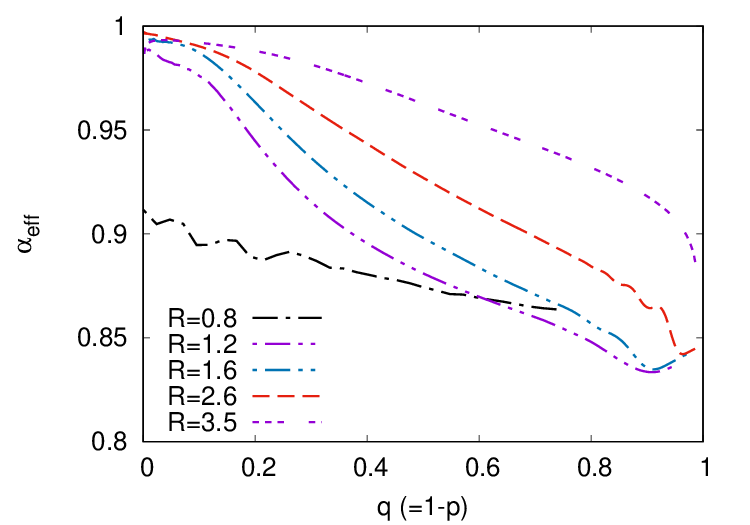}
    \caption{The effective exponent $\alpha_{eff}$ describing the transformed (complementary) Lorenz function for the 2d RFIM is calculated for different values of the disorder parameter. Except for the critical disorder, the effective exponent varies strongly with $q$. For $\delta=1.25$ and the transformation parameter $r=0.05$, the effective exponent for the critical disorder is expected to be around $0.8$, which is close to what is numerically seen. The deviations are due to fluctuations and finite size effects.}
    \label{rfim_aeff}
\end{figure}

Fig. \ref{rfim_avalanche} shows the avalanche size distributions $P(S)\sim S^{-\delta}$ for different values of the disorder strength. The critical strength shows a power law distribution with exponent value close to $-1.25$. Above this disorder strength, the distribution deviates from power law. Below the critical strength, all spins flip in one avalanche. Now, as the critical exponent here is in the range $1<\delta<2$, we have to use the variable transformation sec. II.B. Here we do the transformation $\tilde{S}=S^r$, with $r=0.05$, such that $\tilde{n}=nr<1$, where $n=1/(\delta -1)$. With this transformation, the resulting transformed Lorenz function (particularly, its complement) can then be analyzed. Fig. \ref{rfim_aeff} shows the effective exponent for different values of the disorder strength. The value of the effective exponent, after the transformation with the parameter $r=0.05$, is expected to be $0.8$ for the critical disorder.  The departure from this is due to fluctuations and finite size effects. 

\begin{figure}[tbh]
    \hspace*{-0.5cm}
    \includegraphics[width=1.1\linewidth]{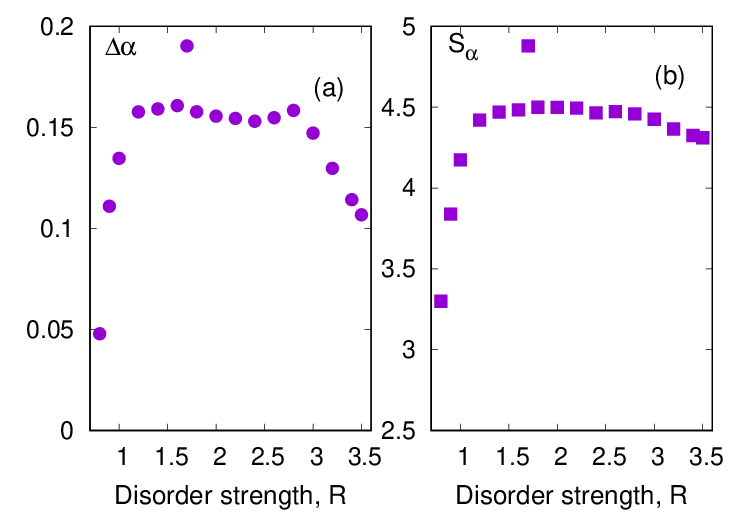}
    \caption{The variations of the range of effective exponent $\Delta\alpha$ and the entropy $S_{\alpha}$ of the distribution of the values of the effective exponent are plotted against the disorder strength. Initially there is a strong correlation, as was seen in the case of synthetic data, before it reaches an intermediate saturation phase around the critical strength.}
    \label{rfim_dalpha}
\end{figure}

As before, we look at the correlation between the indirect estimates of $A(\theta)$, measured through $\Delta\alpha$ and $S_{\alpha}$, and the disorder strength. Fig. \ref{rfim_dalpha} shows the variations of these two quantities with $R$. It can be seen that there is initially a strong correlation between the disorder strength and the two quantities that estimates width of $A(\theta)$. 
It then saturates for an intermediate range of disorder, before finally starting to decay when the avalanches are essentially localized events (for high disorder). 
Therefore, for disorder strength near criticality, these indirect measures of the spectral width can serve as a good indicator of the underlying heterogeneity of the model. 

\section{Discussions and conclusions}
There has been an extensive effort, especially in economics, to estimate the Lorenz function to quantify inequality (see e.g., \cite{econ1,econ2,econ3,econ4}). Under some approximations, the Lorenz function can be fairly accurately inferred, even when the full distribution function (of wealth) is not known. This is a very common scenario in a society, where the complete individual level information of wealth is rarely available, if at all. But the Lorenz function alone can give several useful measures of inequality, for example the Gini and Kolkata indices, that can quantify the wealth concentration in society. The Lorenz function and inequality indices have been also calculated for many other empirical data sets, revealing universal features of inequality statistics \cite{ghosh}.
A more recent application of the Lorenz function and social inequality indices has been to physical systems, particularly when they are near their critical point. For tuned criticality and for SOC, this approach is able to produce universal scaling and precursory signs for imminent large responses in the system (see e.g., \cite{succ}).

In this work, we investigated if the Lorenz function is able to detect any qualitative change in the microscopic heterogeneity of the underlying system. Specifically, a deviation from the underlying power law distribution function could often be due to qualitative shifts in interactions in the microscopic level, for example disorder strength in an avalanching system. In general, microscopic heterogeneity induces local environments with varying effective scaling responses, which collectively appear as a superposition of local exponents. It could also be due to different Pareto tail in wealth distributions of sub-populations. By expanding the Lorenz function in a (non-orthogonal) power-law basis, we are able to extract a correlation between the microscopic heterogeneity (demonstrated here using the random field Ising model) and the spectral width of such an expansion (see Eq. (4)). 

In conclusion, the Lorenz function of a response variable that has a power law size distribution, can detect departures from such a power law in the tail. It has been shown using synthetic data and random field Ising model simulations that such deviations could be related to the underlying microscopic heterogeneity of the system. This could potentially be applied to systems to extract information about its microscopic heterogeneity, where the information for the full distribution function is lacking but the Lorenz curves are well approximated.


\begin{thebibliography}{99}
    \bibitem{lorenz}
    Lorenz, M.O. {\it Methods of Measuring the Concentration of Wealth}. Publ. Am. Stat.
Assoc. 1905, {\bf 9}, 209.
    \bibitem{gini}
    Gini, C. {\it Measurement of Inequality of Incomes}. Econ. J. 1921, {\bf 31}, 124.
    \bibitem{kolkata}
      Ghosh, A.; Chattopadhyay, N.; Chakrabarti, B.K. {\it Inequality in Societies, Academic
Institutions and Science Journals: Gini and k-Indices}. Physica A 2014, 410, 30.
    \bibitem{pareto}
 Pareto, V., A. N. Page Translation of ‘Manuale di economia politica’ (Manual of political economy), A.M. Kelley Publishing, New York (1971).
     \bibitem{neda}
     Biswas, S.; Chakrabarti, B.K.; Ghosh, A.; Ghosh, S.; Józsa, M.; Néda, Z. {\it Does Excellence Correspond to Universal Inequality Level?} Entropy 2025, {\bf 27}, 495.
     \bibitem{ghosh}
     Ghosh, A.; Chakrabarti, B.K. {\it Do Successful Researchers Reach the SelfOrganized Critical Point?} Physics 2024, {\bf 6}, 46–59.
     \bibitem{wool}
     Woolhouse, M.; Dye, C.; Etard, J.; Smith, T.; Charlwood, J.; Garnett, G.; Hagan,
P.; Hii, J.; Ndhlovu, P.; Quinnell, R.; et al. {\it Heterogeneities in the Transmission
of Infectious Agents: Implications for the Design of Control Programs}. Proc. Natl. Acad. Sci. USA 1997, {\bf 94}, 338.
   \bibitem{succ}
      Ghosh, A.; Biswas, S.; Chakrabarti B. K. {\it Success of social inequality measures in predicting critical or failure points in some models of physical systems.}  Front. Phys. 2022, 10:990278. doi: 10.3389/fphy.2022.990278.

  \bibitem{prl}
       Das, S.; Biswas, S. {\it Critical scaling through Gini index},
Phys. Rev. Lett. 2023, {\bf 131}, 157101.
 \bibitem{lomov}    
Lomov, S. V.; Abaimov, S. G.; Breite, C.; Swolfs, Y. {\it Inequality Indices Applied to Statistical Physics of Criticality in an Impregnated Fiber Bundle Model}. Mech. Comp.
Mat. 2023, {\bf 59}, 840.
\bibitem{jordi}
Diksha; Baro, J.; Biswas, S. {\it Inequalities of energy release rates in compression of nano-porous materials predict its imminent breakdown}, Phys. Rev. E, 2024, {\bf 111}, L053502.
 \bibitem{manna}
 Manna, S. S.; Biswas, S.; Chakrabarti, B. K.; {\it Near universal values of social inequality indices in self-organized critical models}, Physica A, 2022, {\bf 596}, 127121.

\bibitem{econ1}
Carr, A. {\it Lorenz interpolation: A method for estimating income inequality from grouped income data}, Sociological Methodology, 2022, {\bf 52}, 141.


\bibitem{econ2}
Jorda, V.; Sarabia, J. M.; Jäntti, M. {\it Inequality Measurement with Grouped Data: Parametric and Non-Parametric Methods}, J. Royal Stat. Soc. Ser. A: Statistics in Society, 2021, {\bf 184}, 964.


\bibitem{econ3}
Sarabia, J. M.; Prieto, F.; Sarabia, M. {\it Revisiting a functional form for the Lorenz curve},  Econ. Lett. 2010, {\bf 107}, 249.
\bibitem{econ4}
 Basmann, R. L.; Hayes, K. J.; Slottje, D. J.; Johnson, J. D. {\it A general functional form for approximating the Lorenz curve}, J. Econometrics, 1990, {\bf 43}, 77.

\bibitem{sethna}
Sethna, J.; Dahmen, K; Myers, C. {\it Crackling noise}, Nature, 2001, {\bf 410}, 242.

\bibitem{newman}
Clauset, A; Shalizi,  C. R.; Newman, M. E. J. {\it Power-Law Distributions in Empirical Data
}, SIAM Reviews, 2009, {\bf 51}, 661.

\bibitem{leg}
Choo, H.; and Ryu, H. {\it Gini coefficient, Lorenz curves, and Lorenz dominance effect: An application to Korean income distribution data.} Journal of Economic Development, 1994, {\bf 19}, 47.

\bibitem{rfim1}
Sethna, J. P.; Dahmen, K.; Kartha, S.; Krumhansl, J. A.; Roberts, B. W.; Shore, J. D. {\it Hysteresis and hierarchies: Dynamics of disorder-driven first-order phase transformations}, Phys. Rev. Lett. 1993, {\bf 70}, 3347.

\bibitem{rfim2}
Perković, O.; Dahmen, K.; Sethna, J. P. {\it Avalanches, Barkhausen Noise, and Plain Old Criticality}, Phys. Rev. Lett. 1995, {\bf 75}, 4528.

\bibitem{rfim3}
Spasojević, D.; Janićević, S.; Knežević, M. {\it Numerical Evidence for Critical Behavior of the Two-Dimensional Nonequilibrium Zero-Temperature Random Field Ising Model}, Phys. Rev. Lett. 2011, {\bf 106}, 175701.

\bibitem{rfim4}
Vives, E.; Rosinberg, M. C.; Tarjus, G. {\it Hysteresis and avalanches in the T=0 random-field Ising model with two-spin-flip dynamics}, Phys. Rev. B, 2005, {\bf 71}, 134424.

\bibitem{rfim5}
Hayden, L. X.; Raju, A.; Sethna, J. P. {\it Unusual scaling for two-dimensional avalanches: Curing the faceting and scaling in the lower critical dimension}, Phys. Rev. Research, 2019, {\bf 1}, 033060.

\bibitem{two_power}
Li, H.; Valdés, E.; Vives, E. {\it Double power-law universal scaling function for the distribution of waiting times in labquake catalogs}, Phys. Rev. E, 2024, {\bf 110}, 064140.

\bibitem{ghosh}
Ghosh, A.; Chattopadhyay, N.; Chakrabarti, B. K. {\it Inequality in societies, academic institutions and science journals: Gini and k-indices}. Physica A, 2014, {\bf 410}, 30.
 
\end{thebibliography}
\end{document}